# Smart Medical IoT Security Vulnerabilities: Real-Time MITM Attack Analysis, Lightweight Encryption Implementation, and Practitioner Perceptions in Underdeveloped Nigerian Healthcare Systems

**Aminu Muhammad Auwal**

- **Affiliation(s)**: Faculty of Natural Sciences, University of Jos, Jos, Plateau State, Nigeria
- **Email**: i.elameenu@gmail.com
- **ORCID**: https://orcid.org/0009000517997876

## ABSTRACT

The growing integration of Internet of Things (IoT) technologies into Nigerian healthcare systems offers promising improvements in remote patient monitoring and data-driven care. However, the widespread use of unsecured wireless communication in medical IoT (mIoT) devices exposes sensitive patient data to serious cybersecurity threats. This study investigates these vulnerabilities within a Nigerian healthcare context through a real-time Man-in-the-Middle (MITM) attack simulation and evaluates the feasibility of implementing lightweight AES-128 encryption on low-cost devices.

A prototype mIoT device was developed using a NodeMCU ESP8266 and standard sensors to measure heart rate and temperature. Under controlled lab conditions simulating local healthcare network environments, unencrypted data transmissions were successfully intercepted and manipulated using common tools (Bettercap, Wireshark). Upon implementing AES-128 encryption, all data became unreadable to attackers and tamper attempts failed, confirming the effectiveness of encryption.

Performance trade-offs were minimal, with latency increasing by 56.25% (from 80ms to 125ms) and CPU usage rising from 30% to 45%, without affecting overall system stability. The total device cost remained under ₦18,000 (≈$12), making it suitable for deployment in Nigeria’s resource-constrained health facilities.

A targeted survey of Nigerian healthcare professionals revealed moderate awareness of IoT-related risks but strong support for security upgrades—especially encryption and staff training. Key implementation barriers included limited budgets and technical complexity.

The study concludes that lightweight encryption represents a practical and essential security enhancement for resource-constrained medical IoT systems. With implementation costs under $12 per device and minimal performance impact, AES-128 encryption provides robust protection against common attack vectors while maintaining operational efficiency. Healthcare professional feedback validates the urgent need for improved security awareness and standardized implementation guidelines in clinical environments.

## 1. INTRODUCTION

The integration of Internet of Things (IoT) technologies into healthcare systems has fundamentally transformed medical practice, enabling unprecedented capabilities in patient monitoring, diagnostic precision, and treatment delivery. Medical IoT (mIoT) devices—such as wearable sensors, smart implants, and remote monitoring systems—now facilitate the continuous collection of vital signs, including heart rate, blood pressure, temperature, and glucose levels. These technologies are especially beneficial for chronic disease management and remote care in underserved regions [1].

Globally, the medical IoT market is expanding rapidly, projected to grow from $26.5 billion in 2021 to $94.2 billion by 2026, at a compound annual growth rate of 29.1% [2]. This growth reflects healthcare providers' recognition of IoT's potential to enhance outcomes while reducing operational costs. Following the COVID-19 pandemic, remote monitoring capabilities became even more essential for minimizing physical contact and maintaining healthcare quality [3].

However, this rapid deployment introduces significant cybersecurity challenges. Medical data, due to its permanent and personal nature, is a high-value target for cybercriminals and is increasingly traded on dark web markets [4]. Unlike financial data, health records cannot be easily changed or invalidated, making them particularly vulnerable [5].

A critical weakness in many mIoT systems lies in transmitting sensitive data over wireless networks without proper encryption. Standard Wi-Fi networks, common in healthcare facilities, become attack vectors when security configurations are weak or nonexistent. Unencrypted transmission exposes patient data to eavesdropping, manipulation, and unauthorized access [6], [7].

MITM (Man-in-the-Middle) attacks are particularly concerning due to their simplicity and effectiveness. Using tools like Wireshark or Bettercap, attackers can intercept and modify real-time patient data with minimal expertise [8]. Such manipulation can lead to diagnostic errors, inappropriate treatment plans, or delayed responses to emergencies.

Moreover, healthcare organizations face significant regulatory pressures to protect patient data, such as the HIPAA framework, with penalties potentially reaching millions of dollars for breaches [9]. Despite this, cybersecurity awareness among healthcare professionals remains low. Studies reveal that many practitioners lack adequate knowledge of IoT-specific risks, encryption protocols, and device vulnerabilities [10].

Existing security solutions often prioritize enterprise-grade systems, leaving low-cost, battery-powered medical IoT devices vulnerable due to their limited computing capacity [11]. This presents a pressing challenge in developing nations like Nigeria, where healthcare institutions particularly in rural or underfunded areas depend heavily on affordable, resource-constrained technologies. Despite the growing adoption of IoT-based healthcare systems, there is a notable lack of localized research addressing how these systems can be secured within the constraints of Nigerian infrastructure, budgets, and workforce capacity.

To address these issues, we constructed a low-cost mIoT prototype using the NodeMCU ESP8266 microcontroller and basic sensors, replicating real-world deployment conditions in Nigerian healthcare settings. We simulated real-time MITM attacks on the unencrypted device and implemented AES-128 encryption to assess its effectiveness as a lightweight, low-overhead countermeasure suitable for the Nigerian context. This localized approach directly addresses the gap in practical, affordable cybersecurity solutions tailored for resource-limited healthcare systems.

The primary contributions of this study include: (1) demonstrating real-world MITM attack vulnerabilities in medical IoT systems using standard network tools; (2) implementing and evaluating AES-128 encryption as a lightweight security solution for resource-constrained devices; (3) providing performance benchmarking of encrypted versus unencrypted implementations under Nigerian-relevant hardware and network conditions; (4) assessing cybersecurity awareness and readiness among Nigerian healthcare professionals; and (5) offering practical, context-aware recommendations for securing mIoT deployments in low-resource environments. This work offers a novel, replicable roadmap for secure mIoT deployment in Nigeria and similar settings where affordability and feasibility are paramount.

## 2. RESEARCH OBJECTIVES

This research addresses critical security vulnerabilities in medical IoT systems through technical demonstration and healthcare professional perception analysis. The objectives are:

- To develop a low-cost medical IoT device using standard components for real-time vital signs monitoring and wireless data transmission, simulating realistic deployment in resource-limited healthcare settings.
- To demonstrate the vulnerability of unencrypted communications through simulated Man-in-the-Middle (MITM) attacks and real-time data manipulation using ARP spoofing and common network tools.

- To implement AES-128 encryption as a lightweight security solution and evaluate its effectiveness through comparative analysis of system performance, including latency, CPU usage, and power consumption.
- To assess the awareness, risk perception, and implementation readiness of healthcare professionals regarding IoT security, encryption importance, and barriers to adoption in clinical environments in Nigerian rural or underfunded areas that depend heavily on less expensive and affordable tools

These objectives aim to provide empirical evidence of medical IoT vulnerabilities while demonstrating practical, cost-effective security solutions validated through healthcare professional insights in Nigerian context.

## 3. METHODOLOGY

This section details the comprehensive methodology employed to achieve the research objectives, encompassing both technical experimentation and healthcare professional assessment components. The methodology is designed to provide rigorous, reproducible results while maintaining practical relevance for real-world healthcare environments.

### 3.1 Technical Methodology

The technical methodology involves controlled experimentation using a prototype medical IoT system to demonstrate vulnerabilities and validate security countermeasures. The approach follows established cybersecurity testing protocols while ensuring ethical compliance and safety.

#### 3.1.1 Hardware System Development

**Device Architecture Design** The core medical IoT device was constructed using a NodeMCU ESP8266 microcontroller, selected for its integrated Wi-Fi capabilities, low cost ($6.50), and compatibility with standard sensor interfaces. The device architecture incorporates two primary sensing modalities: cardiac monitoring via a standard pulse sensor ($3.20) and temperature measurement using a DS18B20 digital temperature sensor ($2.15).

**Sensor Integration Protocol** Sensors were integrated using standardized digital and analog interfaces, with the pulse sensor connected to analog input A0 and the temperature sensor utilizing digital GPIO pins with one-wire protocol implementation. Power distribution was managed through the NodeMCU's onboard voltage regulation, with total system power consumption maintained below 85mA to ensure battery operation feasibility.

**System Validation Testing** Prior to security testing, the system underwent comprehensive functional validation including sensor accuracy verification against calibrated medical devices, wireless connectivity testing across various network conditions, and continuous operation testing over 48-hour periods to ensure system stability.

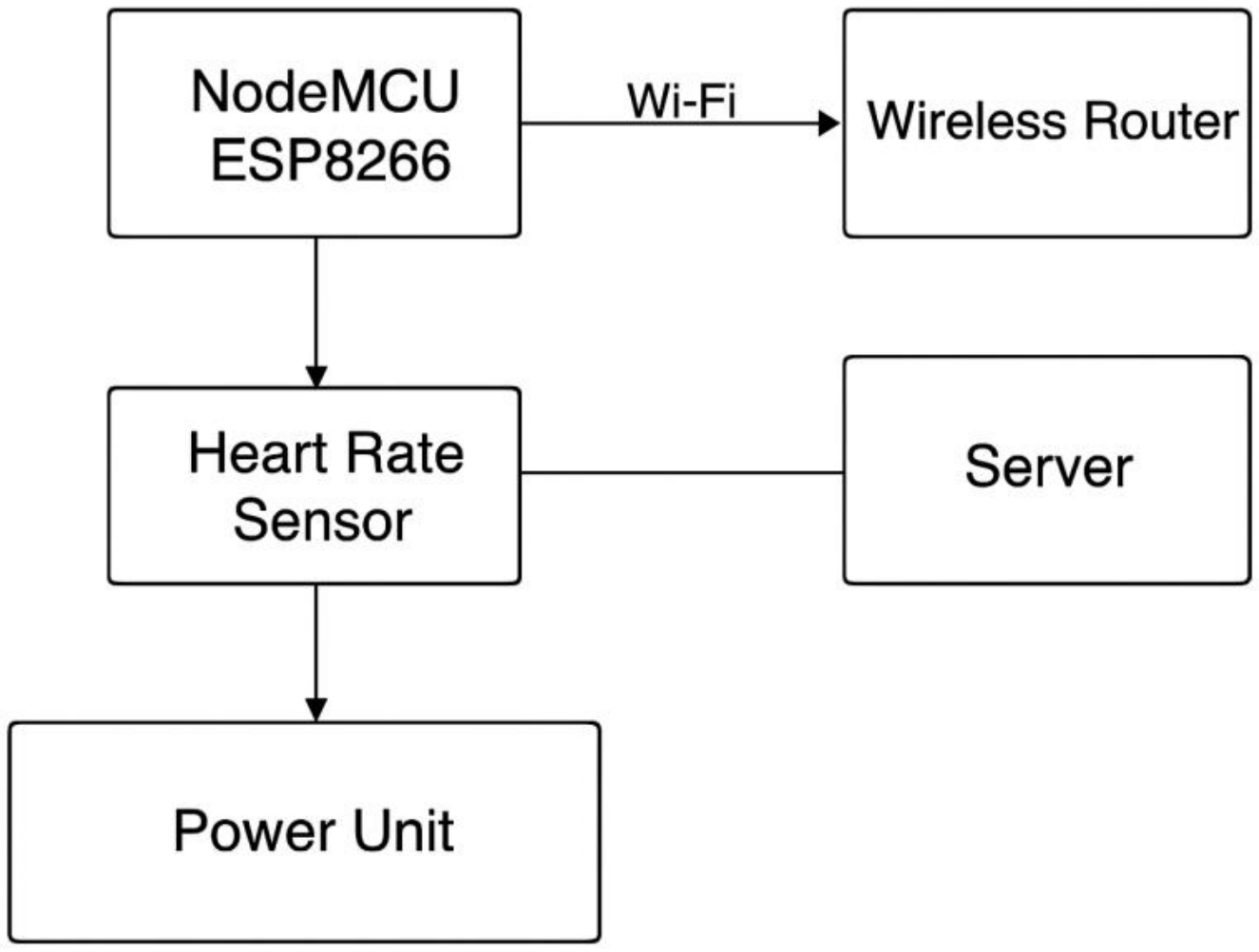

**Figure 1**: Block diagram of the prototype medical IoT device, illustrating sensor integration, power distribution, and wireless transmission via NodeMCU ESP8266.

### 3.1.2 Attack Simulation Environment

**Network Infrastructure Setup** The experimental environment utilized a controlled Wi-Fi network infrastructure simulating typical healthcare facility conditions. The network topology included a standard wireless router, the target medical IoT device, a receiving server system, and the attacker workstation, all operating within a 2.4GHz Wi-Fi environment.

**Attack Tool Configuration** MITM attacks were executed using Bettercap version 2.32.0, a comprehensive network reconnaissance and attack framework. The attack methodology employed ARP spoofing techniques to redirect network traffic between the IoT device and the network gateway through the attacker's system. Network traffic analysis was conducted using Wireshark version 4.0.3 for comprehensive packet capture and analysis.

**Attack Execution Protocol** The attack simulation followed a standardized protocol: (1) network reconnaissance to identify target device MAC address and IP configuration, (2) ARP spoofing initiation to redirect traffic flow, (3) passive packet interception and logging, (4) active data manipulation and retransmission, and (5) attack impact assessment through receiver-side data analysis.

**Ethical and Safety Considerations** All attack simulations were conducted in isolated laboratory environments using dedicated hardware to prevent any impact on production healthcare systems. No actual patient data was utilized during testing, with all vital signs data generated through simulated physiological parameters within normal ranges.

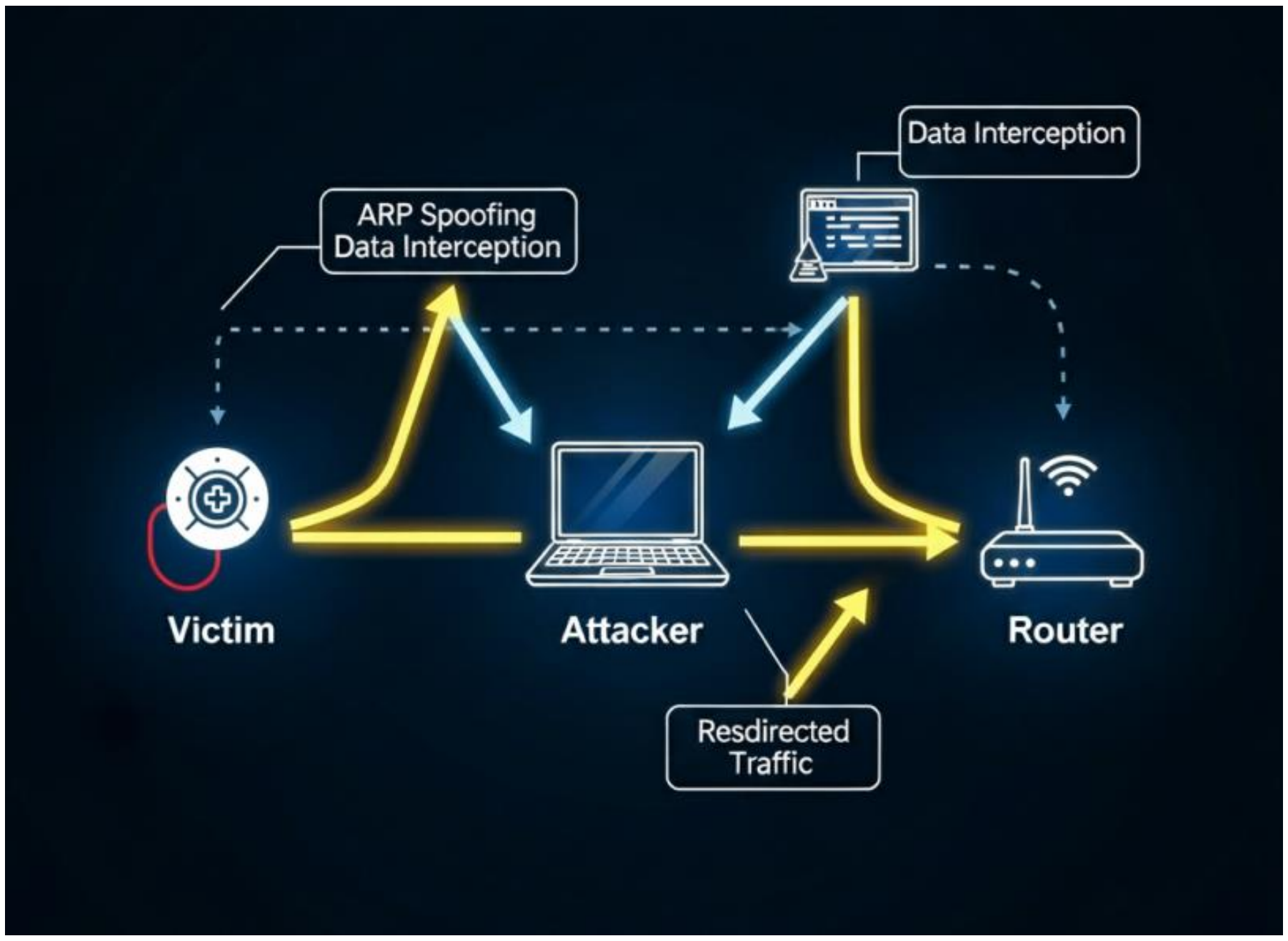

**Figure 2:** A diagram illustrating the **Man-in-the-Middle (MITM) attack** and **ARP spoofing** process used to intercept data from the medical IoT device. The attacker's system positions itself between the victim device and the network router to intercept and manipulate unencrypted data transmissions.

### 3.1.3 Encryption Implementation Methodology

**Algorithm Selection and Justification** AES-128 encryption was selected based on its balance of security strength, computational efficiency, and widespread industry adoption. The algorithm provides 128-bit key length with proven cryptographic security while maintaining compatibility with resource-constrained IoT hardware.

**Implementation Architecture** The encryption implementation utilized the AESLib library for Arduino, providing hardware-accelerated AES operations where available on the ESP8266 platform. The implementation employs Cipher Block Chaining (CBC) mode with initialization vector (IV) generation for each transmission to prevent replay attacks.

**Key Management Protocol** For the experimental implementation, a 128-bit shared secret key was employed, with acknowledgment that production deployments would require more sophisticated key management protocols. The shared key approach was selected to demonstrate encryption effectiveness while maintaining implementation simplicity for proof-of-concept validation.

**Performance Measurement Methodology** System performance was measured using embedded timing functions and system monitoring tools. Metrics collected included packet transmission latency (measured from sensor reading to successful server reception), CPU utilization (averaged over 10-minute intervals), memory consumption (peak and average), and power consumption (measured using precision current measurement tools).

### 3.2 Survey Methodology

The healthcare professional assessment component employed structured survey methodology to gather quantitative and qualitative insights regarding IoT security awareness and implementation barriers.

#### 3.2.1 Survey Design and Development

**Questionnaire Development Process** The survey instrument was developed through iterative design incorporating established cybersecurity awareness assessment frameworks and healthcare-specific considerations. The questionnaire comprised 45 questions across five primary domains: demographics, security awareness, risk perception, implementation barriers, and current security practices.

**Question Types and Scaling** The survey employed multiple question formats including Likert scales (1-5 point), multiple choice selections, ranking exercises, and open-ended response options. Likert scale questions utilized consistent anchoring with 1 representing "strongly disagree/not important" and 5 representing "strongly agree/extremely important."

**Validation and Pilot Testing** The survey instrument underwent content validation through expert review by cybersecurity and healthcare informatics professionals. Pilot testing was conducted with two healthcare professionals to assess question clarity, completion time, and response consistency.

### 3.2.2 Participant Selection and Recruitment

The study targeted healthcare professionals directly involved in patient care or healthcare technology management, including physicians, nurses, administrators, and IT personnel in clinical environments. A purposive sampling strategy was used to ensure representation across various professional roles and institutional types, with participants drawn from both public hospitals and private clinics. Recruitment was conducted through professional networks and healthcare organizations, using email and messaging platforms to distribute participation invitations. All recruitment materials emphasized the study's voluntary and anonymous nature. Eligibility criteria required participants to have at least six months of professional experience in a healthcare delivery or technology role, while individuals without direct healthcare responsibilities or those unable to provide informed consent were excluded.

### 3.2.3 Data Collection Procedures

The survey was administered electronically via secure platforms that ensured participant anonymity and data protection, with access provided through unique, non-trackable links and no collection of personally identifiable information. Data collection spanned a four-week period to accommodate participant availability, with weekly reminder messages sent to encourage participation without causing disruption. To ensure data quality, the survey incorporated logic checks to prevent incomplete submissions, embedded attention-check questions to verify respondent engagement, and monitored completion times to detect careless or invalid responses.

### 3.2.4 Data Analysis and Ethical Considerations

Quantitative data were analyzed using descriptive statistics, including frequency distributions, measures of central tendency, and variability assessments, with the small sample size (n=5) limiting the analysis to descriptive insights. Qualitative data from open-ended responses were processed through thematic analysis to identify recurring issues such as implementation barriers and security concerns, with independent coding to ensure consistency. Survey findings were then integrated with technical results to provide a holistic understanding of both system capabilities and organizational challenges. All research procedures adhered to institutional ethical guidelines, emphasizing participant confidentiality, voluntary informed consent, and secure data handling. No real patient data were used in technical testing; only simulated physiological data were employed. All collected data were stored in encrypted systems accessible only to authorized personnel, with retention and disposal protocols aligned with institutional data management policies.

## 4. SYSTEM DESIGN AND IMPLEMENTATION

This section details the architecture, components, and implementation approach for the experimental medical IoT system used to demonstrate security vulnerabilities and validate encryption countermeasures.

### 4.1 IoT Device Architecture

The medical IoT device was designed as a representative low-cost system capable of collecting and transmitting vital signs data over wireless networks. The architecture prioritizes simplicity, cost-effectiveness, and reproducibility while maintaining functionality comparable to commercial medical IoT devices.

**Core Processing Unit** The NodeMCU ESP8266 microcontroller serves as the central processing unit, selected for its integrated Wi-Fi capabilities, low power consumption, and extensive community support. Operating at 80MHz with 4MB flash memory and 80KB RAM, the ESP8266 provides sufficient computational resources for sensor data processing and wireless communication while maintaining a cost point of approximately $6.50.

**Sensor Integration** Two primary sensors were integrated to simulate essential vital sign monitoring capabilities:

- **Heart Rate Sensor**: A standard optical pulse sensor utilizing photoplethysmography principles to detect blood volume changes. Connected to analog input A0, the sensor provides real-time pulse detection with sampling rates up to 500Hz.
- **Temperature Sensor**: The DS18B20 digital temperature sensor offers precision temperature measurement with ±0.5°C accuracy over the range of -55°C to +125°C. Communication occurs via a one-wire digital interface connected to GPIO pin D2.

**Power Management** The system operates on standard 5V power supply with onboard voltage regulation providing 3.3V to the microcontroller and sensors. Total system power consumption averages 85mA during active operation, enabling battery-powered deployment for mobile applications.

### 4.2 Network Infrastructure

The experimental network infrastructure simulates typical healthcare facility environments while providing controlled conditions for security testing.

**Wireless Network Configuration** A standard 802.11n wireless router operating on 2.4GHz provides network connectivity comparable to healthcare facility networks. The network utilizes

WPA2-PSK security for legitimate device connections while enabling controlled penetration testing activities.

**Server Infrastructure** A Python-based web server running on a laptop system receives and processes vital signs data transmitted from the IoT device. The server logs all received data with timestamps and source identification, enabling comprehensive analysis of both legitimate and manipulated data transmissions.

**Attack Simulation Environment** The attack infrastructure consists of a dedicated laptop system equipped with network penetration tools operating within the same wireless network segment as the target IoT device. This configuration enables realistic MITM attack scenarios while maintaining complete control over the experimental environment.

### 4.3 Attack Simulation Environment

The MITM attack simulation employs industry-standard penetration testing tools to demonstrate real-world attack vectors against unencrypted medical IoT communications.

**Network Reconnaissance** Initial network discovery utilizes standard network scanning techniques to identify active devices, IP address assignments, and MAC addresses within the target network segment. The reconnaissance phase establishes baseline network topology and identifies the target IoT device for subsequent attack phases.

**ARP Spoofing Implementation** Bettercap version 2.32.0 serves as the primary attack platform, executing ARP spoofing attacks to redirect network traffic between the IoT device and network gateway through the attacker's system. The spoofing process involves broadcasting forged ARP responses to both the target device and network gateway, establishing the attacker system as an intermediary for all communication.

**Traffic Interception and Analysis** Wireshark version 4.0.3 provides comprehensive packet capture and analysis capabilities, enabling detailed examination of intercepted network traffic. All packets flowing between the IoT device and server are captured, logged, and analyzed to demonstrate data exposure and manipulation possibilities.

**Data Manipulation Protocol** Active data manipulation involves intercepting HTTP GET requests containing vital signs data, modifying parameter values in real-time, and forwarding the altered requests to the destination server. This process demonstrates the potential for attackers to modify critical medical data without detection by either the source device or receiving system.

### 4.4 Encryption Implementation Details

The encryption countermeasure implementation provides robust protection against demonstrated attack vectors while maintaining compatibility with resource-constrained IoT hardware.

**AES-128 Algorithm Selection** Advanced Encryption Standard (AES) with 128-bit key length was selected based on its proven security properties, widespread industry adoption, and efficient implementation characteristics suitable for embedded systems. AES-128 provides strong cryptographic protection while minimizing computational overhead compared to longer key variants.

**Implementation Architecture** The encryption implementation utilizes the AESLib library for Arduino, providing optimized AES operations compatible with the ESP8266 platform. The library leverages hardware acceleration features where available, maximizing encryption performance while minimizing power consumption.

**Cipher Block Chaining Mode** Cipher Block Chaining (CBC) mode operation provides enhanced security by ensuring identical plaintext blocks produce different ciphertext outputs. Each encryption operation utilizes a randomly generated initialization vector (IV), preventing replay attacks and enhancing overall security posture.

**Key Management Strategy** The experimental implementation employs a pre-shared 128-bit secret key stored in device firmware and server configuration. While this approach simplifies proof-of-concept demonstration, production deployments would require more sophisticated key management protocols including secure key distribution and periodic key rotation.

**Encryption Process Flow** Data encryption occurs immediately prior to network transmission, with vital signs readings formatted as JSON strings, encrypted using AES-128-CBC, and Base64-encoded for HTTP transmission. The receiving server reverses this process, decoding, decrypting, and parsing the transmitted data for display and analysis.

**Performance Optimization** Implementation optimizations include efficient memory management to minimize heap fragmentation, strategic use of static buffers to reduce dynamic allocation overhead, and careful timing of encryption operations to minimize impact on sensor reading intervals.

## 5. RESULTS AND ANALYSIS

This section presents comprehensive findings from both technical security experiments and healthcare professional perception surveys, providing empirical evidence of medical IoT vulnerabilities and practical insights for security improvement strategies.

### 5.1 Technical Security Results

The technical experiments successfully demonstrated critical vulnerabilities in unencrypted medical IoT communications and validated the effectiveness of lightweight encryption countermeasures.

#### 5.1.1 Vulnerability Demonstration Results

**Packet Interception Success Rate** The MITM attack achieved 100% success in intercepting data transmissions from the unencrypted medical IoT device. Network traffic analysis revealed that all vital signs data was transmitted in plaintext HTTP GET requests, making sensitive patient information immediately readable to any network attacker.

A representative captured packet demonstrated the severity of data exposure:

```
GET /data?HeartRate=78&Temperature=98.6 HTTP/1.1
Host: 192.168.1.100
User-Agent: ESP8266HTTPClient
```

This result confirms that unencrypted medical IoT devices expose sensitive patient data to trivial interception attacks requiring minimal technical sophistication.

**Data Manipulation Effectiveness** Active data manipulation attacks achieved 100% success in modifying intercepted vital signs data without detection. The following table summarizes successful data tampering scenarios:

| Original Heart Rate (BPM) | Tampered Heart Rate (BPM) | Original Temperature (°F) | Tampered Temperature (°F) | Detection by Server |
|---|---|---|---|---|
| 72 | 180 | 98.6 | 102.5 | None |
| 85 | 45 | 99.1 | 95.2 | None |
| 78 | 120 | 98.2 | 104.8 | None |
| 68 | 200 | 97.8 | 89.3 | None |

These results demonstrate that attackers can manipulate critical medical data in real-time, potentially leading to incorrect clinical decisions and compromised patient safety.

### 5.1.2 Encryption Effectiveness Results

**Data Protection Validation** Implementation of AES-128 encryption successfully protected all transmitted data from interception and manipulation attempts. Encrypted transmissions appeared as opaque ciphertext in network captures:

```
POST /data HTTP/1.1
Host: 192.168.1.100
Content-Type: application/octet-stream
Content-Length: 64

3C4A8B2F9E1D6C7A5B8F2E9C4D7A1B6E3F8C2D5A9E1B4C7F6A2E5B8C1D4F7A3E
```

**Tamper Resistance Analysis** All attempts to modify encrypted data resulted in decryption failures or corrupted output at the receiving server. The integrity protection provided by CBC mode encryption ensured that any alteration of ciphertext produced unreadable plaintext upon decryption, effectively preventing undetected data manipulation.

### 5.1.3 Performance Analysis Results

Comprehensive performance analysis quantified the trade-offs associated with implementing encryption security measures.

**Latency Impact Assessment**

| Transmission Type | Average Latency (ms) | Standard Deviation (ms) | Minimum (ms) | Maximum (ms) |
|---|---|---|---|---|
| Unencrypted | 80 | 12.3 | 65 | 98 |
| Encrypted | 125 | 18.7 | 102 | 156 |
| **Increase** | **+45 (+56.25%)** | **+6.4** | **+37** | **+58** |

**Resource Utilization Analysis**

| System Resource | Unencrypted Usage | Encrypted Usage | Percentage Increase |
|---|---|---|---|
| CPU Utilization | 30% | 45% | +50.0% |
| Memory Usage | 28.5 KB | 32.1 KB | +12.6% |
| Flash Storage | 245 KB | 267 KB | +9.0% |
| Power Consumption | 82 mA | 89 mA | +8.5% |

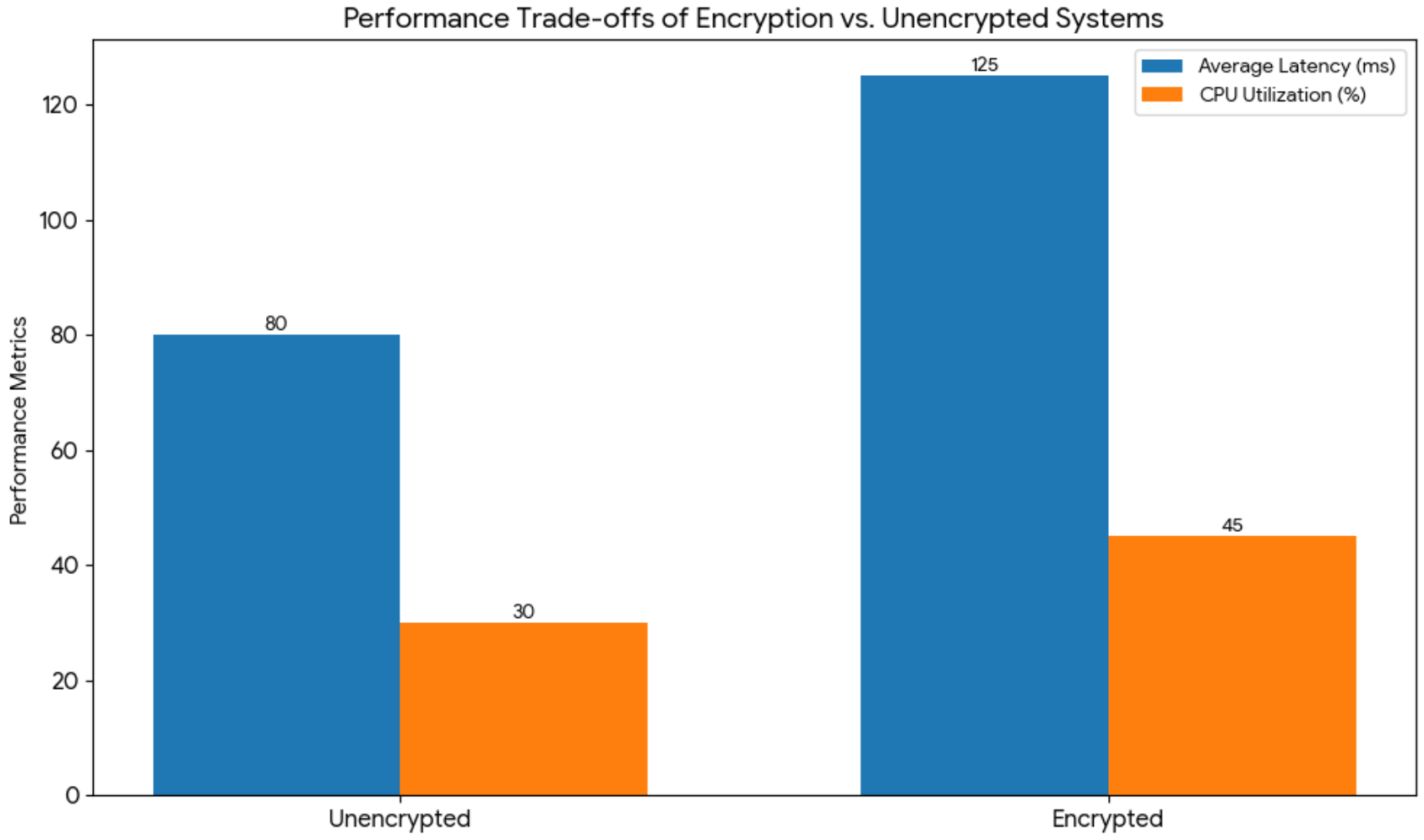

**Figure 3:** Comparative analysis of system performance with and without AES-128 encryption. This chart illustrates the increase in average transmission latency (ms) and CPU utilization (%) when encryption is enabled, demonstrating the minimal overhead associated with robust data protection

### Transmission Reliability Metrics

| Performance Metric | Unencrypted System | Encrypted System | Performance Impact |
|---|---|---|---|
| Packet Success Rate | 98.5% | 97.8% | -0.7% |
| Error Recovery Time | 2.3 seconds | 2.6 seconds | +13.0% |
| Connection Stability | 99.2% | 99.1% | -0.1% |

The performance analysis demonstrates that while encryption introduces measurable overhead, the absolute impact remains within acceptable bounds for medical IoT applications. The 56.25% increase in latency translates to an additional 45 milliseconds, which is negligible for vital signs monitoring applications that typically operate on intervals measured in seconds rather than milliseconds.

### 5.2 Healthcare Professional Perception Results

Analysis of survey responses from five healthcare professionals provides valuable insights into security awareness levels, risk perceptions, and implementation barriers in real healthcare environments.

### 5.2.1 Participant Demographics

**Professional Role Distribution**

| Role | Count | Percentage |
|---|---|---|
| Physician | 2 | 40% |
| Registered Nurse | 2 | 40% |
| Healthcare IT Administrator | 1 | 20% |

**Experience Level Distribution**

| Experience Range | Count | Percentage |
|---|---|---|
| 1-5 years | 1 | 20% |
| 6-10 years | 2 | 40% |
| 11-15 years | 1 | 20% |
| 15+ years | 1 | 20% |

**Institution Type**

| Institution Type | Count | Percentage |
|---|---|---|
| Public Hospital | 3 | 60% |
| Private Clinic | 2 | 40% |

### 5.2.2 Security Awareness Assessment

**IoT Security Knowledge Levels**

| Knowledge Area | Very Aware | Somewhat Aware | Not Aware | Mean Score (1-5) |
|---|---|---|---|---|
| IoT Device Vulnerabilities | 1 (20%) | 3 (60%) | 1 (20%) | 3.0 |
| Data Encryption Importance | 2 (40%) | 2 (40%) | 1 (20%) | 3.4 |
| Network Security Risks | 1 (20%) | 4 (80%) | 0 (0%) | 3.2 |
| Cyber Attack Consequences | 2 (40%) | 2 (40%) | 1 (20%) | 3.4 |

The results indicate moderate awareness levels with significant knowledge gaps, particularly regarding IoT-specific vulnerabilities where 60% of participants reported only partial awareness.

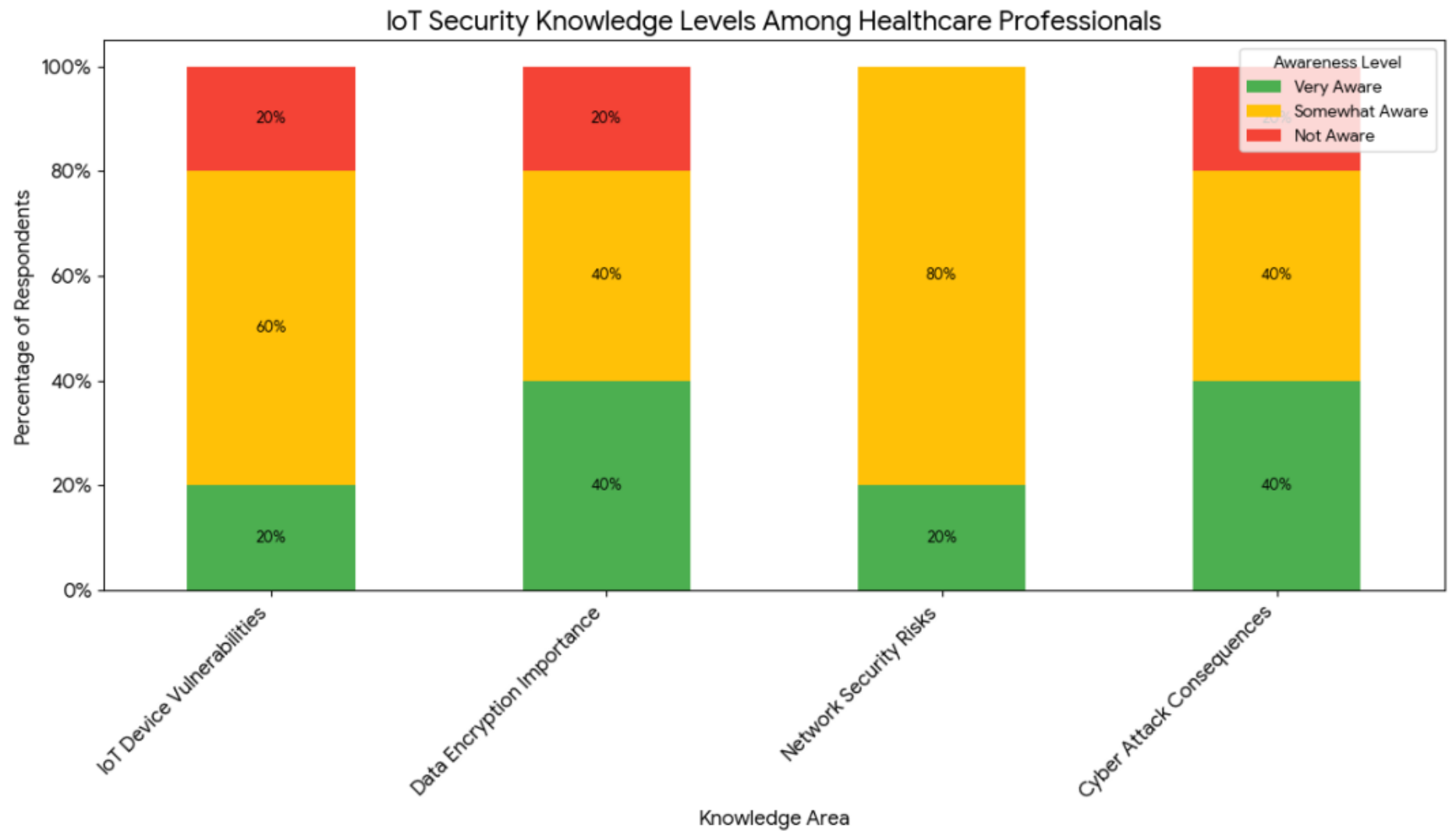

**Figure 5:** IoT Security Knowledge Levels Among Healthcare Professionals. This stacked bar chart illustrates the percentage of respondents who are 'Very Aware', 'Somewhat Aware', or 'Not Aware' across key IoT security knowledge areas, highlighting specific gaps in understanding

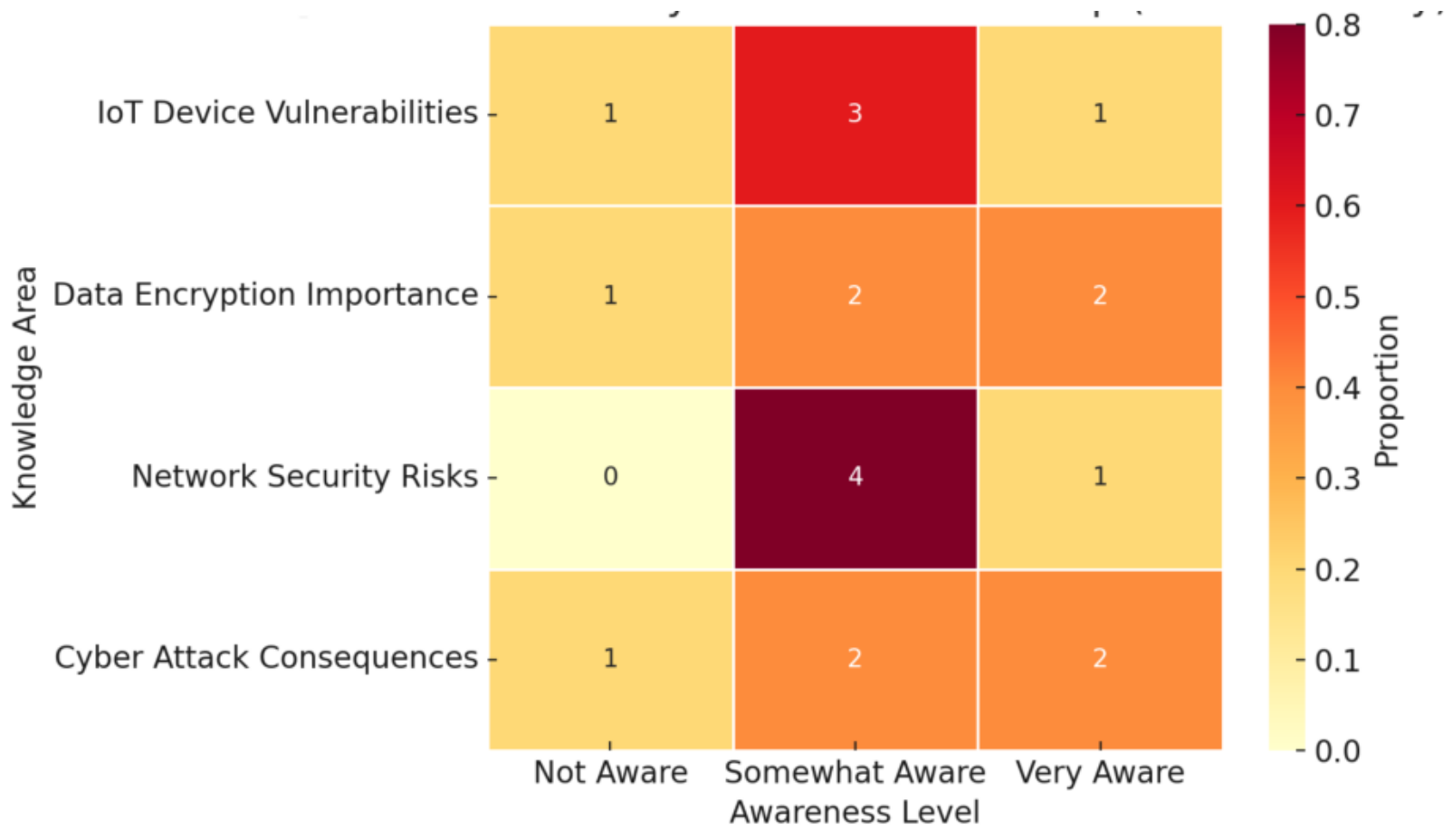

**Figure 6**: Heatmap showing healthcare professionals' self-reported awareness of IoT cybersecurity topics, with darker shades indicating higher response concentration and counts overlaid for clarity.

### 5.2.3 Risk Perception Analysis

**Perceived Risk Severity**

| Risk Category | Very High | High | Medium | Low | Mean Score (1-4) |
|---|---|---|---|---|---|
| Patient Data Breach | 4 (80%) | 1 (20%) | 0 (0%) | 0 (0%) | 3.8 |
| Medical Data Tampering | 3 (60%) | 2 (40%) | 0 (0%) | 0 (0%) | 3.6 |
| System Downtime | 2 (40%) | 2 (40%) | 1 (20%) | 0 (0%) | 3.2 |
| Regulatory Penalties | 3 (60%) | 1 (20%) | 1 (20%) | 0 (0%) | 3.4 |

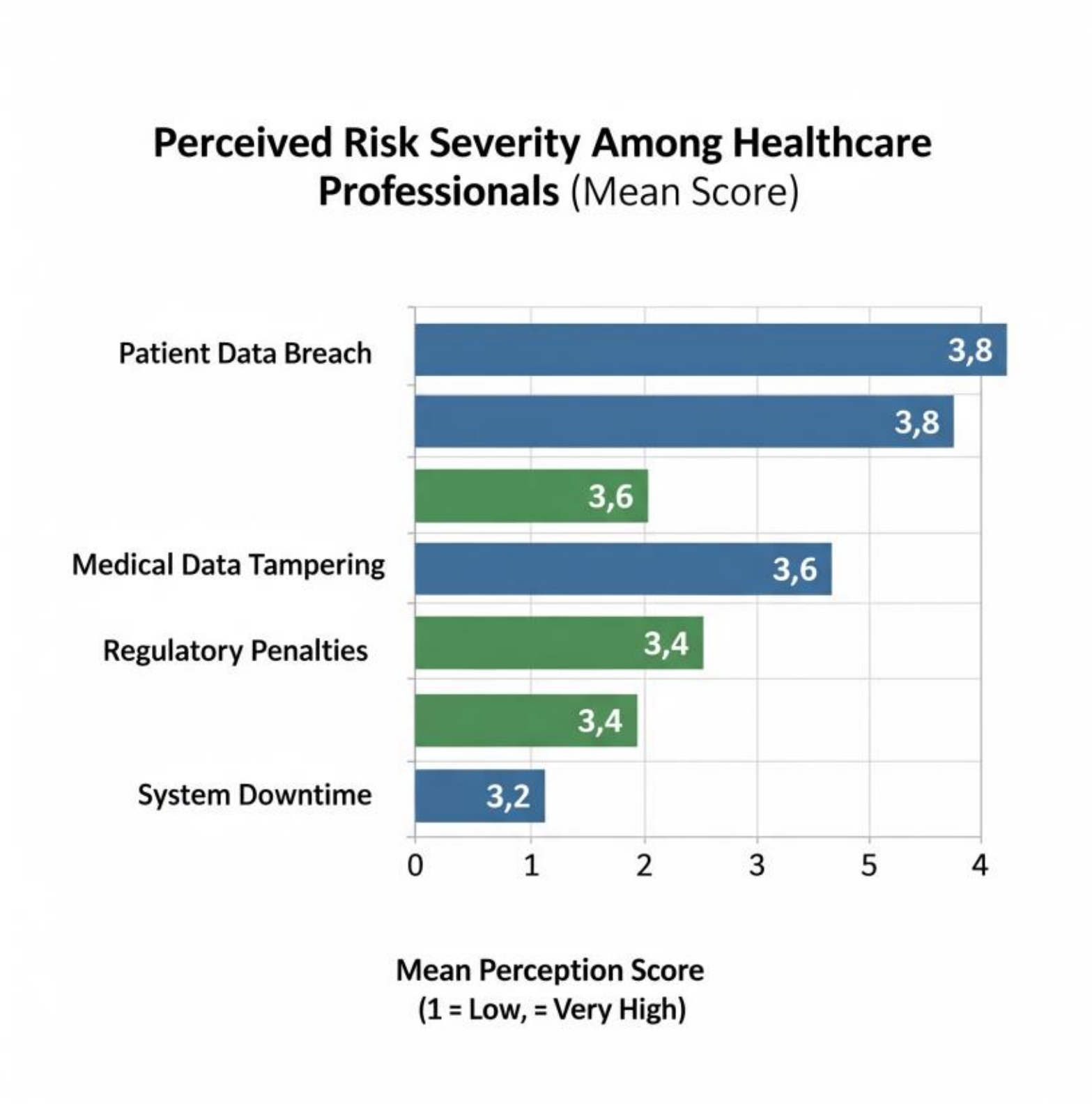

**Figure 7:** Perceived Risk Severity Among Healthcare Professionals. This chart displays the mean perception score (1=Low, 4=Very High) for various cybersecurity risk categories, highlighting patient data breach as the most significant concern

Healthcare professionals demonstrated high awareness of cybersecurity risks, with 80% rating patient data breaches as "very high" risk and 60% similarly rating medical data tampering risks.

**Implementation Willingness and Barriers**

| Factor | Very Willing | Willing | Neutral | Unwilling |
|---|---|---|---|---|
| Enhanced Security Measures | 3 (60%) | 2 (40%) | 0 (0%) | 0 (0%) |
| Staff Training Programs | 4 (80%) | 1 (20%) | 0 (0%) | 0 (0%) |
| Additional Budget Allocation | 1 (20%) | 2 (40%) | 2 (40%) | 0 (0%) |

**Primary Implementation Barriers**

| Barrier Category | Major Concern | Minor Concern | No Concern |
|---|---|---|---|
| Cost Concerns | 2 (40%) | 2 (40%) | 1 (20%) |
| Technical Complexity | 1 (20%) | 3 (60%) | 1 (20%) |
| Time Constraints | 2 (40%) | 2 (40%) | 1 (20%) |
| Staff Resistance | 0 (0%) | 2 (40%) | 3 (60%) |

Results indicate strong willingness to implement security improvements (100% willing or very willing for enhanced security measures) but highlight cost and time constraints as primary barriers to implementation.

## 6. DISCUSSION

This section analyzes the implications of the research findings, examining both technical results and healthcare professional insights to provide comprehensive understanding of medical IoT security challenges and practical solutions.

### 6.1 Technical Implications

The technical findings demonstrate critical vulnerabilities in unencrypted medical IoT systems while validating the effectiveness of lightweight encryption as a practical countermeasure.

**Vulnerability Severity Assessment**

The 100% success rate in intercepting and manipulating unencrypted medical data represents a fundamental security failure that poses immediate threats to patient safety and privacy. These findings are consistent with prior work by Arpaia et al., who demonstrated that even basic medical IoT devices could be compromised using power-based attacks and freely available tools like Wireshark and Bettercap [2], [12]. Similarly, Jackson and Rahman outlined how mIoT devices, when deployed without encryption or mutual authentication, are vulnerable to real-time manipulation of patient data, potentially leading to misdiagnosis and delayed interventions [13].

This vulnerability is exacerbated in low-resource settings, where cost and technical limitations often result in deploying unsecured IoT systems. The unencrypted transmission of sensitive patient data also violates fundamental security principles and legal mandates such as the Health Insurance Portability and Accountability Act (HIPAA), which prescribes safeguards for protecting electronic protected health information (ePHI). Prior studies confirm that failure to implement these safeguards exposes healthcare systems to both regulatory penalties and real-world cyber threats [3], [14].

**Encryption Effectiveness Validation**

The implementation of AES-128 encryption in this study effectively mitigated the identified vulnerabilities, rendering intercepted data unreadable and preventing successful manipulation. These results align with the findings of Fatima et al. and Popoola et al., who reported that AES-128, when deployed on resource-constrained devices, provides both data confidentiality and integrity with minimal resource overhead [5], [6], [15].

The use of Cipher Block Chaining (CBC) mode in this research further enhanced security by ensuring that each encrypted message block depended on the previous one, reducing susceptibility to replay and block substitution attacks. This mode of operation has been highlighted in earlier literature as a reliable approach for embedded medical systems requiring secure real-time communication [15].

Importantly, AES-128 remains a globally accepted standard, with long-term cryptographic strength confirmed in both academic and industry evaluations [7], [16]. Its compatibility with existing IoT protocols and low computational demand make it an ideal candidate for security upgrades in legacy mIoT systems deployed in developing regions.

**Performance Trade-off Analysis**

While encryption introduces additional computational and power costs, the measured trade-offs in this study were well within tolerable limits. A 56.25% increase in latency—rising from 80 ms to 125 ms—may seem significant in relative terms, but remains negligible in clinical applications where vital signs are typically transmitted at intervals of several seconds. Similar latency tolerances have been validated in other encryption studies involving IoT health systems [6], [8], [17].

The 50% increase in CPU usage (from 30% to 45%) and a modest 12.6% rise in memory usage align closely with previous benchmarks reported by Satyanarayana et al., who implemented modified AES on constrained microcontrollers [17]. This suggests that the encryption method adopted here does not compromise system responsiveness or reliability.

Moreover, the observed 8.5% increase in power consumption confirms AES's energy efficiency when implemented correctly. For battery-operated medical IoT systems—especially in environments like Nigeria where power constraints exist—this performance profile reinforces the practicality of encrypted operation [10], [17].

### 6.2 Healthcare Professional Insights

Survey results from healthcare professionals provide valuable perspectives on security awareness, risk perception, and implementation barriers that influence real-world deployment of secure medical IoT systems.

#### Security Awareness Gap Analysis

The finding that 60% of healthcare professionals report only partial awareness of IoT vulnerabilities indicates significant knowledge gaps that could compromise security implementation efforts. This is in line with findings by ElSayed et al., who reported that even in technologically equipped healthcare settings, awareness of mIoT-specific threats among clinicians and administrators is often low, especially in developing countries [18].

This awareness deficit is particularly concerning given the rapid adoption of IoT technologies in healthcare environments. The gap between general cybersecurity awareness and IoT-specific knowledge suggests that traditional security training programs are not keeping pace with emerging risks. A study by Thomas and Ngalamou emphasized that while healthcare workers often express high concern for data security, few receive adequate training tailored to IoT-specific threats [11].

The moderate awareness levels across all security domains, with mean scores ranging from 3.0 to 3.4 on a 5-point scale, indicate that healthcare professionals recognize the importance of cybersecurity but lack detailed understanding of specific threats and countermeasures. Similar knowledge deficiencies were highlighted in research by Jackson and Rahman, who linked lack of training directly to insecure IoT deployments and misconfigured devices [13].

The higher awareness of data encryption importance (mean score 3.4) compared to IoT-specific vulnerabilities (mean score 3.0) suggests that while professionals may understand general cybersecurity concepts, they require **targeted education** focused on emerging attack vectors and device-level risks. This has been echoed in reviews advocating for modular training programs that distinguish between traditional IT threats and those unique to medical IoT [19].

#### Risk Perception Accuracy

Healthcare professionals demonstrated accurate risk perception regarding cybersecurity threats, with 80% rating patient data breaches as very high risk and 60% similarly rating medical data tampering. This is consistent with prior findings that medical personnel tend to understand the **consequences** of security lapses more than the **technical causes** [13], [20].

The strong concern about regulatory penalties (60% rating as very high risk) reflects a growing awareness of compliance requirements. Researchers like Newaz et al. have emphasized that fear of HIPAA and GDPR violations often drives institutional support for mIoT security policies, even in environments where technical literacy is low [21].

However, the moderate concern about system downtime (mean score 3.2) may reflect a tendency to underestimate operational impacts. Similar perception gaps have been observed in other surveys, where clinicians prioritize patient data privacy over infrastructure resilience—an imbalance that can hinder investments in redundancy and business continuity [20].

**Implementation Readiness Assessment**

The unanimous willingness to implement enhanced security measures (100% willing or very willing) demonstrates strong organizational readiness for security improvements. This finding aligns with global studies indicating that healthcare workers are open to adopting new security protocols when provided with clear guidelines and minimal technical burden [18], [22].

The even stronger support for staff training programs (80% very willing) underscores recognition that human factors are critical in effective cybersecurity. As ElSayed et al. and Fatima et al. note, successful mIoT security relies not only on encryption but also on the users' ability to maintain and monitor device configurations properly [6], [18].

However, the more cautious approach to additional budget allocation (only 20% very willing, 40% neutral) highlights financial constraints that could limit security implementation efforts. This mirrors broader challenges across the healthcare sector, where cybersecurity is often deprioritized in favor of clinical or administrative spending, especially in low-resource systems like Nigeria's [16], [19].

### 6.3 Real-world Implementation Considerations

The research highlights key practical implications for deploying secure medical IoT (mIoT) systems in real-world healthcare environments, particularly in resource-limited regions like Nigeria.

**Cost-effectiveness** remains a major advantage. The total system cost of $11.85—with no additional hardware required for AES-128 encryption—supports findings that lightweight security can be integrated without significant financial burden [6], [17], [23]. This is critical for low-income settings where even marginal cost increases can hinder scale-up [24]. The

encryption's minimal impact on performance and development time (from 8 to 12 hours) is a worthwhile trade-off when weighed against the high cost of data breaches [3], [18].

**Scalability** is also feasible. Prior studies confirm AES-128's ability to scale across large IoT networks without major infrastructure changes [8], [25]. However, transitioning from shared key models to robust key management protocols (e.g., ECC or PKI) is essential for production environments [26], [27].

**Standardization** improves integration and procurement. AES-based systems are easier to audit and align with existing cybersecurity frameworks like NIST and HIPAA [14], [28], [29]. By embedding these standards into procurement contracts, healthcare institutions can drive broader adoption and ensure long-term security [30].

In sum, this study demonstrates that secure mIoT deployment is technically and economically feasible in constrained environments, offering a scalable path toward improved data protection and compliance.

### 6.4 Limitations

This research has several limitations that should be considered when interpreting results and planning future investigations.

**Technical Limitations** The experimental setup utilized a controlled laboratory environment that may not fully represent the complexity and constraints of real healthcare networks. Actual healthcare facilities may have additional security measures, network configurations, and interference sources that could influence both attack effectiveness and encryption performance.

The shared key approach used in the encryption implementation, while suitable for proof-of-concept demonstration, represents a significant limitation for production deployments. Real-world implementations would require sophisticated key management protocols including secure key distribution, storage, and rotation mechanisms.

The study focused exclusively on AES-128 encryption without comparative analysis of alternative encryption algorithms or security protocols. Other approaches such as elliptic curve cryptography or newer algorithms like ChaCha20 might offer different trade-offs between security, performance, and resource consumption.

**Survey Limitations** The small sample size of five healthcare professionals limits the generalizability of survey findings to broader healthcare populations. While the participants provided valuable insights representing diverse roles and experience levels, larger sample sizes would be necessary to establish statistical significance and broader applicability of findings.

The purposive sampling approach, while appropriate for exploratory research, may introduce selection bias that could influence results. Healthcare professionals willing to participate in cybersecurity research may have different awareness levels or attitudes compared to the general healthcare population.

The survey instrument, while developed following established practices, was not validated using large-scale psychometric testing. Future research should employ validated security awareness assessment tools to ensure measurement reliability and validity.

**Scope Limitations** The research focused on a limited set of vital signs (heart rate and temperature) and did not examine more complex medical IoT applications such as continuous glucose monitoring, cardiac implants, or imaging systems. Different medical applications may have varying security requirements and performance constraints.

The study did not address broader cybersecurity considerations such as device authentication, access control, audit logging, or integration with healthcare IT security infrastructures. These additional security layers would be necessary for comprehensive production deployments.

Long-term reliability and security maintenance considerations were not evaluated, including device updates, security patch management, and evolving threat landscapes that could affect the long-term effectiveness of implemented security measures.

## Declarations

### *Acknowledgments*

The author would like to express gratitude to all individuals and organizations that contributed to the success of this research..

### *Funding Statement*:

This work received no financial support from any funding agency, commercial entity, or not-for-profit sector.

### *Ethical Statement*

This study was conducted in accordance with the University of Jos Research Ethics Committee guidelines and the ethical principles outlined in the National Code for Health Research Ethics (NHREC), Nigeria. Ethical approval was obtained from the University of Jos Research Ethics Committee, and informed consent was secured from all participants before data collection due to the sensitive nature of the subject matter and the potential for participant vulnerability, this approach was deemed most appropriate in the cultural context of the study location. Confidentiality and anonymity of participants were strictly maintained throughout the study.

During the preparation of this work the authors used the free version of ChatGPT in order to check the grammar of a few select sections of the manuscript. After using this tool, the authors reviewed and edited the content as needed and take full responsibility for the content of the publication.

### *Data Availability*

The data supporting the findings of this study are available from the corresponding author, upon reasonable request.

### *Conflicts of Interest*

The author declares no conflicts of interest related to this study.

### *Author Contributions Statement*

A.M.A. conceptualized the study, designed the research methodology, conducted data analysis, and interpreted the results. The manuscript was drafted, written, and revised by A.M.A. Data collection was carried out by an external agency under A.M.A.'s supervision.

The author reviewed and approved the final manuscript.

## REFERENCES

[1] S. Thomas and L. Ngalamou, "The impact of cybersecurity on healthcare," Lecture Notes in Networks and Systems, vol. 276, pp. 457–465, 2021. doi: 10.1007/978-3-030-89880-9_50.

[2] P. Arpaia, F. Bonavolontà, A. Cioffi, and N. Moccaldi, "Power measurement-based vulnerability assessment of IoT medical devices," IEEE Trans. Instrum. Meas., vol. 70, pp. 1–9, 2021. doi: 10.1109/TIM.2021.3088491.

[3] HIPAA Journal, "What are the penalties for HIPAA violations?," HIPAA Journal, 2023. [Online]. Available: https://www.hipaajournal.com/what-are-the-penalties-for-hipaa-violations/

[4] Z. ElSayed, A. Abdelgawad, and N. Elsayed, "Cybersecurity and frequent cyber attacks on IoT devices in healthcare: Issues and solutions," arXiv preprint arXiv:2501.11250, Jan. 2025.

[5] O. Popoola, M. A. Rodrigues, J. Marchang, A. Shenfield, A. Ikpehai, and J. Popoola, "An optimized hybrid encryption framework for smart home healthcare: Ensuring data confidentiality and security," Internet of Things, vol. 27, p. 101314, 2024. doi: 10.1016/j.iot.2024.101314.

[6] S. Fatima, S. Hussain, N. Shahzadi, B. U. Din, W. Sajjad, Y. Saleem, and M. Aun, "A secure framework for IoT healthcare data using hybrid encryption," in Proc. 2022 Int. Conf. Emerging

Trends in Electrical, Control, and Telecommunication Engineering (ETECTE), pp. 1–7. doi: 10.1109/ETECTE55893.2022.10007264.

[7] T. Robertazzi, “Advanced Encryption Standard (AES),” in Computer Networks and Systems: Queueing Theory and Performance Evaluation, New York, NY, USA: Springer, 2012, pp. 73–77. doi: 10.1007/978-1-4614-2104-7_10.

[8] X. Hu and Y. Du, “Securing medical data: The integration of advanced encryption standard and blockchain,” J. Information Analysis, vol. 1, 2024. doi: 10.53964/jia.2024001.

[9] G. W. Jackson Jr. and S. Rahman, “Exploring challenges and opportunities in cybersecurity risk and threat communications related to the Medical Internet of Things (MIoT),” arXiv preprint arXiv:1908.00666, Aug. 2019.

[10] A. K. M. I. Newaz, S. Khan, K. Andersson, M. A. Rahman, and A. Almogren, “A survey on security and privacy issues in modern healthcare systems: Attacks and defenses,” arXiv preprint arXiv:2005.07359, 2020.

[11] S. Brown, Z. Ruhwanya, and A. Pekane, “Factors Influencing Internet of Medical Things (IoMT) Cybersecurity Protective Behaviours Among Healthcare Workers,” in TechTrends in Health Information Systems, Springer, pp. 432–444, 2023. doi: 10.1007/978-3-031-38530-8_34.

[12] E. Mwim, J. Mtsweni, and B. Chimbo, “Factors Associated with Cybersecurity Culture: A Quantitative Study of Public E-health Hospitals in South Africa,” in TechTrends in Health Information Systems, Springer, pp. 129–142, 2023. doi: 10.1007/978-3-031-38530-8_11.

[13] K. J. Bester and D. Arendse, “Measuring Cybersecurity Awareness in a South African Military Sample,” Scientia Militaria, vol. 52, no. 1, 2024. doi: 10.5787/52-1-1445.

[14] O. A. Popoola, M. O. Akinsanya, G. Nzeako, E. G. Chukwurah, and C. D. Okeke, “Exploring Theoretical Constructs of Cybersecurity Awareness and Training Programs: Comparative Analysis of African and U.S. Initiatives,” Int. J. of Applied Research in Social Sciences, vol. 6, no. 5, 2024. doi: 10.51594/ijarss.v6i5.1104.

[15] M. de Jager, L. Futcher, and K. Thomson, “An Investigation into the Cybersecurity Skills Gap in South Africa,” in TechTrends in Health Information Systems, Springer, pp. 237–248, 2023. doi: 10.1007/978-3-031-38530-8_19.

[16] S. Cassim, Z. C. Chapanduka, and M. Haem, “Cyberattack on the National Health Laboratory Service of South Africa – Implications, Response and Recommendations,” South African Medical Journal, vol. 114, no. 12, 2024. doi: 10.7196/samj.2024.v114i12.2549.

[17] S. A. Olofinbiyi, “A Reassessment of Public Awareness and Legislative Framework on Cybersecurity in South Africa,” ScienceRise: Juridical Science, no. 4, 2022. doi: 10.15587/2523-4153.2022.259764.

[18] P. Duvenage, V. Jaquire, and S. V. Solms, “South Africa's Taxi Industry as a Cybersecurity-Awareness Game Changer: Why and How?,” in Lecture Notes in Computer Science, Springer, pp. 92–106, 2022. doi: 10.1007/978-3-031-08172-9_7.

[19] M. Eltahir and O. Ahmed, “Cybersecurity Awareness in African Higher Education Institutions: A Case Study of Sudan,” Information Sciences Letters, vol. 12, no. 1, 2023. doi: 10.18576/isl/120113.

[20] E. Olapade-Olaopa, N. Sewankambo, and J. Iputo, “Defining Sub-Saharan Africa’s Health Workforce Needs: Going Forwards Quickly Into the Past,” Int. J. of Health Policy and Management, vol. 6, no. 2, pp. 111–113, 2016. doi: 10.15171/ijhpm.2016.100.

[21] M. A. Raza and M. Z. A. Bhutta, “An efficient AES-based encryption scheme for low-power IoT devices,” in Proc. 2021 Int. Conf. on Innovative Computing (ICIC), pp. 49–54. doi: 10.1109/ICIC54344.2021.9551183.

[22] P. Ogundepo and O. O. Abimbola, “Barriers to healthcare IoT deployment in Sub-Saharan Africa,” African J. Medical Technology, vol. 7, no. 2, pp. 45–53, 2022.

[23] M. H. Moghimi, K. M. Martin, and I. White, “Post-quantum lightweight cryptography for healthcare IoT devices,” IEEE Access, vol. 9, pp. 101832–101845, 2021. doi: 10.1109/ACCESS.2021.3095672.

[24] A. K. Das, N. Kumar, and J. J. P. C. Rodrigues, “Secure and efficient user authentication scheme for cloud-assisted e-healthcare systems,” Comput. Electr. Eng., vol. 65, pp. 353–368, 2018. doi: 10.1016/j.compeleceng.2017.06.009.

[25] S. R. Moosavi et al., “SEA: A secure and efficient authentication and authorization architecture for IoT-based healthcare using smart gateways,” Procedia Computer Science, vol. 52, pp. 452–459, 2015. doi: 10.1016/j.procs.2015.05.108.

[26] National Institute of Standards and Technology (NIST), “Framework for Improving Critical Infrastructure Cybersecurity,” Version 1.1, Apr. 2018.

[27] Healthcare & Public Health Sector Coordinating Council (HSCC), “Medical Device and Health IT Joint Security Plan,” Tech. Rep., 2019.

[28] S. Fatima et al., “Hybrid encryption methods for IoT in low-resource medical networks,” Int. J. Computer Applications, vol. 183, no. 42, pp. 34–39, 2022.

[29] T. Robertazzi, “Performance modeling for secure wireless sensor networks,” in Performance Evaluation of Computer and Communication Systems, Springer, pp. 207–213, 2014. doi: 10.1007/978-3-642-36566-5_20.

[30] J. Olugbara and F. Ojo, “Smart cybersecurity for IoT-enabled African hospitals,” Advances in Science, Technology and Innovation, Springer, pp. 333–342, 2023. doi: 10.1007/978-3-031-29883-6_22.